\documentstyle[prd,aps]{revtex}
\newcommand{\nc}{\newcommand}
\nc{\beq}{\begin{equation}}   \nc{\eeq}{\end{equation}}
\nc{\bea}{\begin{eqnarray}}   \nc{\eea}{\end{eqnarray}}
\nc{\baa}{\begin{array}}      \nc{\eaa}{\end{array}}
\nc{\bit}{\begin{itemize}}    \nc{\eit}{\end{itemize}}
\nc{\ben}{\begin{enumerate}}  \nc{\een}{\end{enumerate}}
\nc{\bce}{\begin{center}}     \nc{\ece}{\end{center}}
\def\beqa{\begin{eqnarray}}
\def\eeqa{\end{eqnarray}}

\def\MPL #1 #2 #3 {{\sl Mod. Phys. Lett.}~{\bf#1} (#3) #2}
\def\NPB #1 #2 #3 {{\sl Nucl. Phys.}~{\bf #1} (#3) #2}
\def\PLB #1 #2 #3 {{\sl Phys. Lett.}~{\bf #1} (#3) #2}
\def\PR #1 #2 #3 {{\sl Phys. Rep.}~{\bf#1} (#3) #2}
\def\PRD #1 #2 #3 {{\sl Phys. Rev.}~{\bf #1} (#3) #2}
\def\PRL #1 #2 #3 {{\sl Phys. Rev.~Lett.}~{\bf#1} (#3) #2}
\def\RMP #1 #2 #3 {{\sl Rev. Mod. Phys.}~{\bf#1} (#3) #2} 
\def\ZPC #1 #2 #3 {{\sl Z. Phys.}~{\bf #1} (#3) #2}
\def\IJMP #1 #2 #3 {{\sl Int. J. Mod. Phys.}~{\bf#1} (#3) #2}
\def\NIM #1 #2 #3 {{\sl Nucl. Inst. and Meth.}~{\bf#1} {#3} #2}

\def\mx{M_X}
\def\delm{\Delta m}
\def\sigmin{\sigma_{\rm min}}
\def\sigsm{\sigma_{\rm SM}}
\def\phiz{\phi_Z^0}
\def\mz{m_Z}
\def\ki{K_i}
\def\kiform{|\langle \phiz|h_i^0\rangle|^2}
\def\hi{h_i^0}
\def\mhi{m_{\hi}}

\def\rts{\sqrt s}

\def\h{h}
\def\mh{m_{\h}}
\def\mhmax{\mh^{\rm max}}
\def\mhmin{\mh^{\rm min}}

\def\eg{{\it e.g.}}

\def\epem{e^+e^-}
\def\mupmum{\mu^+\mu^-}

\def\lsim{\mathrel{\raise.3ex\hbox{$<$\kern-.75em\lower1ex\hbox{$\sim$}}}}
\def\gsim{\mathrel{\raise.3ex\hbox{$>$\kern-.75em\lower1ex\hbox{$\sim$}}}}
\def\@versim#1#2{\vcenter{\offinterlineskip
        \ialign{$\m@th#1\hfil##\hfil$\crcr#2\crcr\sim\crcr } }}

\def\ie{{\it i.e.}}

\def\fbi{~{\rm fb}^{-1}}
\def\fb{~{\rm fb}}

\def\gev{\,{\rm GeV}}

\def\hsm{h_{\rm SM}}

\def\mz{m_Z}

\def\h{h}
\def\mh{m_{\h}}

\def\ie{{\it i.e.}}

\def\9{\phantom 0}      
\renewcommand\linebreak{\unskip\break} 
\newlength{\captsize} \let\captsize=\small 

\def\simlt{\stackrel{<}{{}_\sim}}

\newcommand\be{\beq}
\newcommand\ee{\eeq}
\begin{document}
\draft
\input epsf
\def\la{\mathrel{\mathpalette\fun <}}
\def\ga{\mathrel{\mathpalette\fun >}}
\def\fun#1#2{\lower3.6pt\vbox{\baselineskip0pt\lineskip.9pt
        \ialign{$\mathsurround=0pt#1\hfill##\hfil$\crcr#2\crcr\sim\crcr}}}

\twocolumn[\hsize\textwidth\columnwidth\hsize\csname
@twocolumnfalse\endcsname

\title{A No-Lose Theorem for Higgs Searches at a Future Linear Collider}
\author{Jose R. Espinosa$^{(1)}$ and John F. Gunion$^{(2)}$ }
\address{\phantom{ll}}
\address{
$^{(1)}${\it CERN TH-Division, CH-1211 Geneva 23, Switzerland}\\
$^{(2)}${\it University of California, Davis CA 95616, USA}
}
\date{\today}
\maketitle
\begin{abstract}
Assuming perturbativity up
to a high energy scale $\sim 10^{16}$ GeV, we demonstrate that a future
$e^+e^-$ linear collider operating at $\sqrt{s} =$ 500 GeV with
$\int{\cal L}=$ 500 fb$^{-1}$ per year (such as the recently proposed TESLA
facility) will detect a Higgs boson signal
regardless of the complexity of the Higgs sector
and of how the Higgs bosons decay.
\end{abstract}
\pacs{PACS: 14.80.Cp, 12.60.Jv, 12.20.Fv   \hskip 5.5 cm\\
CERN-TH/98-225~~  UCD-98-12~~ IEM-FT-176/98~~ \hskip 1.5 cm
hep-ph/9807275}
\vskip2pc]

{\bf 1.} In this paper, we show that an $\epem$ collider of sufficient
luminosity (which we quantify) will be guaranteed to find a signal for a
Higgs boson \cite{hhg} if the model remains perturbative up to energy
scales of order $10^{16}\gev$. The maximal luminosity is
that needed to detect a broad enhancement in the $\mx$ spectrum
in $\epem\to Z X\to \epem X+\mupmum X$ over the perturbatively 
allowed $\mx$ mass region. 

For simplicity, we will adopt a notation that assumes CP conservation
in the Higgs sector; generalization is straightforward.
A general neutral Higgs field 
$\Phi_\alpha$ in an arbitrary $(t_\alpha,Y_\alpha)$ representation
is written in terms of its real (CP-even) and imaginary (CP-odd)
components as
$\Phi^0_\alpha=
\frac{1}{\sqrt{2}}(h_\alpha^{0r}+v_\alpha+ih_\alpha^{0i})\,,$
where $v_{\alpha}$ is the vacuum expectation value acquired
by the field as a result of electroweak symmetry breaking.

The crucial ingredient for the no-lose theorem is that there must be Higgs
bosons with mass below and in the vicinity of a perturbatively limited
mass scale $m_B$ (of order the Fermi scale) which, in aggregate,
have net coupling-squared to $ZZ$ that is at least as large as
the SM value. The proof begins with two crucial inequalities.
Defining 
\be
\phiz=\frac{1}{N_Z}\sum_\alpha t_{3\alpha}^2h_\alpha^{0r}v_\alpha\,,
\ee
where $N_Z$ is a normalization constant, the first inequality is \cite{LW,CE}
\be
\label{phizeq}
\langle \phiz|M^2|\phiz\rangle \leq \lambda v^2\equiv m_B^2\,,
\label{msqineq}
\ee
where $M^2$ is the mass-squared operator for the neutral Higgs bosons, 
$\lambda$ is some adimensional quartic Higgs coupling and
$v^2=1/(\sqrt{2} G_F)\sim (246\ {\mathrm GeV})^2$.
The coupling $\lambda$ is not asymptotically free and can be bounded above by
triviality arguments assuming it remains perturbative up to a high-energy
scale $\Lambda$. 
The model-dependence of $m_B$
will be discussed below. The second inequality is
\be
g^2_{ZZ\phiz}\geq g^2_{ZZ\hsm}=g^2\mz^2/\cos^2\theta_W\,.
\label{gsqz}
\ee
(The equality holds only for $t_{3\alpha}=0,\pm 1/2$.)
We now define $\ki=\kiform$, where
$\sum_i \ki=1$,
and insert complete sets of $\hi$ states in 
(\ref{msqineq}) to obtain 
\beq
\langle \phiz|M^2|\phiz \rangle=
\sum_i \ki\mhi^2\leq m_B^2\,.
\label{msqlim}
\eeq
Further, since $\phiz$ defines the combination of states that couples
to $ZZ$, we have $g^2_{ZZ\hi}=\ki g^2_{ZZ\phiz}\,.$
Similar results apply in the $WW$ sector.
 
{\bf 2.} In the SM with  $\Lambda\sim M_{Pl}$, $m_B\sim 190$ GeV
\cite{SM}, and modest luminosity at $\sqrt{s}=500$ GeV would be sufficient 
to detect the unique Higgs boson \cite{lc}. In the MSSM, there are two 
CP-even Higgs bosons, and, for
$\Lambda_{SUSY}\leq 1$ TeV and $\Lambda=\Lambda_{GUT}\sim 10^{16}$ GeV,
$m_B\simlt 125$ GeV \cite{MSSM}.
Eq.~(\ref{msqlim}) implies that, when $K_1\rightarrow 0$, $m_{h_2}\leq
m_B$ and $K_2\rightarrow 1$. For $\sqrt{s}$ well above $m_B$, this is 
sufficient to guarantee Higgs detection in the $\h_{1,2} Z$ channels
alone\cite{lc}. (At
LEP2, the complementary channel $\h_1 A$ must be used to cover the
$K_1\rightarrow 0$ case although large radiative corrections render that
channel inoperative in some small region of parameter space \cite{hole}.)
In the NMSSM (the MSSM extended
by a singlet chiral multiplet), $m_B\sim 140$ GeV \cite{NMSSM} for the same
$\Lambda_{GUT}$, and there are three CP-even Higgs bosons. 
Using the fact that $\sigma_{Z\h}(\mh)$
is a decreasing function of $\mh$ (for constant $g^2_{ZZ\h}$) and using
our previous results, the minimum cross section
for any one of the Higgs bosons,
\be
\sigmin\equiv {\rm min}\left[\left\{\rm max\right\}_i 
\sigma_{Z\h}(\hi)\right]\,,
\label{sigmin}
\ee
[where $\sigma_{Z\h}(\hi)=\ki\sigsm(\mhi)$]
must be bigger than \cite{KOT,NMSSM} $\sigsm(m_B)/3$, as achieved for 
$\mhi=m_B$ and $\ki=1/3$ for $i=1,2,3$.
For $\sqrt{s}=300$ GeV (and $m_B=140$ GeV), $\sigmin\sim 61$ fb.
(This improves over Ref.~\cite{KOT} by a different use of the mass
inequalities.) For this $\sigmin$, an integrated luminosity of
$L=10\fbi$ would be more than adequate to detect at least
one of the
three CP-even scalars. In contrast, there are substantial regions
of NMSSM parameter space for which none of the Higgs bosons will be
discovered at the LHC \cite{lhc}.

In general SUSY
models with arbitrary particle content,
$m_B$ can be as high as 200 GeV for $\Lambda\sim 10^{17}$
GeV \cite{eq98}. As fundamental bosons at the electroweak scale are
natural only in SUSY, we will take $m_B\simlt 200$ GeV for the general
case. Experimentally, the success of fits to precision
electroweak data using a single light Higgs boson
with SM-like couplings and mass $\lsim 200\gev$ implies that,
in a multi-Higgs model, the Higgs bosons with large $ZZ$
coupling must have average mass $\lsim 200\gev$.

{\bf 3.} The minimal cross section (\ref{sigmin}) gets smaller if more
singlets
are added to the model \cite{KOT} (a large number of singlets is
predicted in some string models) or if there is CP violation with
mixing of scalar and pseudoscalar Higgses. Roughly,
$\sigmin\sim\sigma_{SM}(m_B)/N$ where $N$ counts the number of neutral
mass eigenstates that can mix with $\phiz$.
However, for large $N$
there is inadequate mass resolution to separate the various Higgs
bosons if they all have mass $\sim m_B$.  One should simply start
adding their signals together (in all our discussions,
we assume that the widths of the individual Higgs boson eigenstates
are sufficiently small that interference effects can be neglected)
inside mass bins of size determined by the experimental resolution,
$\delm$ (defined as full width at half maximum). 
Consequently, the worst-case situation in the limit
of large $N$ arises when all of the Higgs bosons that can couple to
$\phiz$ have mass spread out below and somewhat above $m_B$,
but are spaced more closely than $\delm$,
and have coupling $\ki$ such that $\ki\sigsm(\mhi)$
closely mimics the background of relevance. 
In this way, the only observable becomes the excess of overall
rate as compared to that expected from background over
some mass interval. The resonance peaks of the Higgs bosons will
not be observable.  

To analyze this situation more quantitatively, we 
first note that once the spacing
is smaller than $\delm$ it becomes irrelevant as to how
many Higgs bosons there are.  Thus, we will simply convert
to an integral notation. Our crucial sum rules become:
\bea
&\int& d\mh K(\mh)=1 \label{ksumint}\\
&\int& d\mh K(\mh)\mh^2\leq m_B^2\,.
\label{msqlimint}
\eea
Since it seems overly perverse to assume that $K(\mh)\sigsm(\mh)$
is exactly proportional to the background cross section as a function
of $\mh$ (after cuts and efficiencies),
we assume a spectrum of Higgs bosons that is uniform
in $\mh$ up to some
maximum value $\mhmax$ starting from $\mhmin$. 
For $K(\mh)=K$, (\ref{ksumint}) implies $K=1/(\mhmax-\mhmin)$,
and (\ref{msqlimint}) implies 
\bea
m_B^2&\geq&
{1\over 3}\left([\mhmax]^2+\mhmax\mhmin+[\mhmin]^2\right)\,.
\label{newlim}
\eea
The maximal spread is achieved for $\mhmin=0$, in which case
(\ref{newlim}) requires $\mhmax\leq \sqrt 3 m_B$.
(The minimal spread, arising when $\mhmax=\mhmin\leq m_B$,
is not of interest for our worst-case considerations.)
To have the greatest sensitivity to the maximal spread region, we 
ideally require $\rts$ large enough that $\sigsm(\mh)$ is large
for all $\mh\lsim \sqrt 3m_B$.  
For $m_B=200\gev$, $\sqrt 3m_B\sim 340\gev$ and quite large $\rts$
would be required to achieve this ideal.  However,
$\rts\sim 500\gev$ would not be too bad. For this $\rts$, 
$\sigsm(\mh)$ varies from $\sim 70\fb$ at low $\mh$
falling to $\sim 15\fb$ at $\mh\sim 340\gev$. At $\mh\sim 200\gev$,
$\sigsm(\mh)$ has only fallen to $\sim 42\fb$. For this reason, 
we will confine ourselves to the $\leq 200\gev$
window in $\mh$ in discussing experimental sensitivity. 

Another important issue is the channel(s)
in which we must detect the Higgs bosons. In the worst
case, the Higgs bosons will decay invisibly (\eg\ to two LSP's
in a supersymmetric model) or to a large number of channels. 
In these cases, identification of the final state would either not
be possible or would not be useful because of the large background
in any one channel (respectively).  The only robust procedure
is to look for $\epem\to Z\h$ and, possibly, $\epem\to\epem\h$ ($ZZ$-fusion)
production in the 
$\epem\to Z X$ (with $Z\to\epem$ and $\mupmum$) and $\epem\to \epem X$
channels, respectively --- we employ cuts
that separate $\epem\to Z\h,Z\to\epem$ from $\epem\to\epem\h$ via
$ZZ$-fusion --- 
as an excess in the recoil $\mx$ distribution due to the continuum
of Higgs bosons. Our focus will be on the $Z\h$ channel. 

{\bf 4.} We find it most convenient to use the analysis of
Ref.~\cite{ghs}.
Although the cuts employed there restrict the analysis to
the region $\mh=70-200\gev$, the results will be very representative
of what can be achieved. Roughly, a fraction $130\gev/340\gev\sim 0.4$
of the signal weight would be in this region
for $\mhmin=0$ and $\mh$-independent $K$. Of course, roughly
1/5 of this worst-case continuum Higgs spectrum would lie 
in the $\lsim 70\gev$ region
to which LEP2 certainly has sensitivity.
We believe that more weight in this region can
probably already be eliminated using current
data. If the LEP limits prove even stronger, then we would have to increase
$\mhmin$ to an appropriate lower limit and our conclusions for
a future $\epem$ machine would be improved.
For example, for $\mhmin=70\gev$ (\ie\ very small $K$ being required
by LEP2 data below this), then $\mhmax\sim 300\gev$ is required
for $m_B=200\gev$. and a fraction $130/230\sim 0.55$ of the Higgs
continuum signal would be in the mass window we shall consider.

The results of Ref.~\cite{ghs} are best summarized in two tables.
In Table~\ref{ghsrates}, we give the approximate signal and background rates
for $L=500\fbi$ for the $Z\h$ and $ZZ$-fusion contributions
to the $\epem\to\epem \h$ channel, and the corresponding $S/\sqrt B$
values obtained by integrating over two different mass intervals:
$70-200\gev$, which includes the $Z$ peak in the background; and
$100-200\gev$, which starts above the $Z$ peak.
For both intervals we have assumed a ``continuum'' of Higgs bosons
in the mass interval, the sum of whose coupling-squared to $ZZ$
is equal to that of the SM Higgs boson.  
The $Z\h$ with $Z\to\mupmum$ channel would give $S$ and $B$ values
that are essentially the same as for $Z\h$ with $Z\to\epem$.
Summing the $\epem$ and $\mupmum$ $Z\h$ channels yields $S\sim 1350$,
and $S/\sqrt B\sim 17$ and $\sim 26$ for the $70-200\gev$
and $100-200\gev$ windows in $\mx$, respectively.
Of course, both $S$ and $S/\sqrt B$ must be reduced by the
fraction $f$ of the Higgs signal
contained in the mass interval. Above, we argued
that $f\sim 0.4 - 0.55$ is an appropriate range to consider. 
Then, for $L=500\fbi$,
we would have a net signal of $S\sim 1350 f$ with a background
of either $B=6350$ or $B=2700$, depending on which of the two
mass intervals are considered. For the $70-200\gev$ ($100-200\gev$) interval,
we have to detect the presence of a broad $\sim 21\%f$ 
($\sim 50\%f$) excess over background, respectively.
Understanding the absolute normalization of the backgrounds
to the $20\%f$ level of accuracy is probably feasible at an $\epem$ collider
for $f\gsim 0.4$; a $50\%f$ excess would certainly be observable. 

We have included the $\epem\to\epem \h$ $ZZ$-fusion channel 
in Table~\ref{ghsrates} to indicate how difficult its use would be
in this situation. Even though $ZZ$-fusion yields a large
nominal $S/\sqrt B$, the $S/B$ values are $0.032f$ and $0.058f$
for the $70-200\gev$ and $100-200\gev$ intervals, respectively.
The first excess would almost certainly be impossible to detect
as a broad enhancement.  Even the $\sim 6\%$ ($f=1$) excess over the latter
interval would be difficult to be certain of.

If we decrease $L$ to $50\fbi$,
we lose a factor of $\sqrt{10}$ in our $S/\sqrt B$ values,
leaving effects at the $S/\sqrt B=5.3f$ and $8.2f$ level
in the $Z\h$ (with $Z\to\epem+\mupmum$) channel for
the $70-200\gev$ and $100-200\gev$ intervals, respectively, which 
is borderline for $f\sim 0.4-0.55$.

{\bf 5.} If such a broad continuum-like signal is observed, the next
important
step will be to divide the $\mx$ spectrum up
into bins, knowing that in each bin there is presumably an
excess of events due to some set of Higgs bosons with some
portion of the net SM $ZZ$ coupling. Just to illustrate
what the situation might be, suppose we distribute the 
roughly 1350 SM signal events (for $Z\h$ with $Z\to\epem$ and $\mupmum$)
equally in the thirteen 10 GeV bins from 70 to 200 GeV.
The level of signal and background, and corresponding $S/\sqrt B$ value,
in each bin is given in Table~\ref{ghsbin}. Note that both
$S$ and $S/\sqrt B$ would more generally be reduced by the fraction $f$
of the SM signal present in the $70-200\gev$ mass window.

Table~\ref{ghsbin} shows that $L=500\fbi$ would 
yield $S/\sqrt B > 3$ only for the $\mx\gsim 120\gev$ bins
when $f\sim 0.5$. With only $L=100\fbi$ (as might be
achieved after a few years of running at a lower luminosity design),
this bin-by-bin type of analysis would 
not be possible for 10 GeV bins if $f\sim 0.5$.

It is useful to determine the number of Higgs bosons that would necessitate
considering the continuum limit.
Suppose there were $13$ spread out over the
$70\gev$ to $200\gev$ interval, \ie\ a Higgs boson every $10\gev$.
The resulting signals would be impossible
to resolve into separate mass peaks if $\delm\gsim 10\gev$.
In fact, this is nearly twice as
good as the typical $\mx$ resolution found 
in Ref.~\cite{ghs} using `super--JLC' tracking (which, in
turn, is substantially
better than currently planned for first-round NLC detectors).
Thus, for current detector designs
it would appear that a model with $\gsim 10$ Higgs bosons 
that can couple to $ZZ$
would be in danger of leading to a continuum Higgs signal.

{\bf 6.} In summary, we have considered a `worst-case' Higgs 
scenario with a large number of Higgs bosons which
cannot be separated from one another given expected
detector resolutions, and which
all decay invisibly or to many final states. In this case, the only 
viable signal is a continuum
excess in the $\epem\to Z X$ (with $Z\to\epem$ and $\mupmum$) recoil
$\mx$ spectrum. We have shown
that the assumption of perturbativity up to high
scales and the bounds/inequalities derived therefrom,
guarantee the observability of this signal
provided that an integrated luminosity in excess of $L=100\fbi$
is accumulated. Bin-by-bin analysis of such a continuum excess would very
possibly require $L\geq 500\fbi$, implying that
this situation provides a rather strong
motivation for going to a very high $L$ collider design.

{\it Acknowledgements.} JRE thanks Y.~Okada and P.~Zerwas for helpful 
discussions. JFG is supported by the U.S. DOE and by DIHEP.

\begin{table}[h]
\caption[fake]{Approximate $S$, $B$ and 
$S/\sqrt B$ values for the $Z\h$ 
and $ZZ$-fusion contributions to $\epem\to\epem \h$, after integrating
the $\mx$ recoil mass spectrum 
from (a) 70 GeV to 200 GeV and (b) 100 GeV to 200 GeV, assuming
that many Higgs bosons are distributed evenly
throughout the interval and have $\sum_{\h}g_{ZZ\h}^2=g_{ZZ\hsm}^2$.
Results are for $\rts=500\gev$, $L=500\fbi$,
and assume the cuts and efficiencies of Ref.~\cite{ghs}.}
\begin{center}
\small
\begin{tabular}{|c|ccc|ccc|}
\hline
 $\mx$  & \multicolumn{3}{c|} {$Z\h$} & \multicolumn{3}{c|} {$ZZ$-fusion} \\
 Interval  & $S$ & $B$ & ${S/\sqrt B}$ & $S$ & $B$ & ${S/\sqrt B}$ \\
\hline
$70-200$ & 675 & 3170 & 12.0 & 2615 & 80100 & 9.2 \\
$100-200$ & 678 & 1350 & 18.5 & 2300 & 39600 & 11.5 \\
\hline
\end{tabular}
\end{center}
\label{ghsrates}
\end{table}

\begin{table}[h]
\caption[fake]{Approximate $S$, $B$ and 
$S/\sqrt B$ values for $Z\h$ (with $Z\to\epem+\mupmum$) 
in each of the thirteen 10 GeV bins in $\mx$
from 70 to 200 GeV, assuming that $S\sim 1350$ events,
from Higgs bosons with $\sum_{\h}g_{ZZ\h}^2=g_{ZZ\hsm}^2$,
are distributed equally among these bins.
We assume $\rts=500\gev$, $L=500\fbi$, and use the results
of Ref.~\cite{ghs}.}
\begin{center}
\small
\begin{tabular}{|c|c|c|c|c|c|c|c|c|}
\hline
Bin No. & 1 & 2 & 3 & 4 & 5 & 6 & 7 & 8--13 \\
$S$ & 104 & 104 & 104 & 104 & 104 & 104 & 104 & 104 \\
$B$ & 1020 & 1560 & 1440 & 734 & 296 & 162 & 125 & $\sim 130$ \\
$S/\sqrt B$ & 3.3 & 2.6 & 2.7 & 3.8 & 6.0 & 8.2 & 9.3 & 9.1 \\
\hline
\end{tabular}
\end{center}
\label{ghsbin}
\end{table}

\hfill\eject
\end{document}